\documentclass[a4paper, 11pt]{article}
\usepackage{amsmath}
\usepackage{graphicx}
\usepackage{latexsym}

\def\b{\begin{eqnarray}}
\def\e{\end{eqnarray}}
\def\n{\noindent}

\begin{document}

\begin{center}

{\LARGE\textbf{Conformal and geometric properties of the
Camassa-Holm hierarchy \\}} \vspace {10mm} \vspace{1mm} \noindent

{\large \bf Rossen Ivanov}\footnote{Present  address: School of
Mathematical Sciences; Dublin Institute of Technology, Kevin
Street, Dublin 8, Ireland, E-mail: rivanov@dit.ie} \vskip1cm \n
\hskip-.3cm
\begin{tabular}{c}
\hskip-1cm $\phantom{R^R}${\it School of Mathematics, Trinity
College,}
\\ {\it Dublin 2, Ireland} \\
\\ {\it E-mail: ivanovr@maths.tcd.ie} \\
\\
\hskip-.8cm
\end{tabular}
\vskip1cm
\end{center}


\begin{abstract}

\n Integrable equations with second order Lax pair like KdV and
Camassa-Holm (CH) exhibit interesting conformal properties and can
be written in terms of the so-called conformal invariants (Schwarz
form). These properties for the CH hierarchy are discussed in this
contribution.

The squared eigenfunctions of the spectral problem, associated to
the Camassa-Holm equation represent a complete basis of functions,
which helps to describe the Inverse Scattering Transform (IST) for
the Camassa-Holm hierarchy as a Generalised Fourier Transform
(GFT). Using GFT we describe explicitly some members of the CH
hierarchy, including integrable deformations for the CH equation.
Also we show that solutions of some 2+1-dimensional
generalizations of CH can be constructed via the IST for the CH
hierarchy.



\end{abstract}


\section{Introduction}

\n Integrable equations exhibit many extraordinary features, like
infinitely many conservation laws, multi- Hamiltonian structures,
soliton solutions etc.  Many integrable equations in {1+1}
dimensions like KdV, MKdV, Harry-Dym, Boussinesq equations possess
interesting conformal properties as well
\cite{D03,D87,W88,N89,DS95,Lou}. These properties originate from
the associated spectral problem, which in most of the cases is
related to a second order differential operator.

The Camassa-Holm (CH) \cite{CH93} equation, which became famous as
a model in water-wave theory \cite{J02,J03,DGH03,DGH04}, together
with its complete integrability
\cite{CH93,FF81,C01,CM99,CI06,CGI,CGI2} exhibits the same type of
conformal properties as well \cite{PLA05}.

The origin of these properties can be understood noticing that for
a second order Lax operator $L=\partial^2+f(x)$, the Poisson
structure is generated by the operator \cite{D03} \b
L^{3/2}_{+}=\partial^3+ \frac{3}{4}(f\partial+\partial\circ f). \e

The CH equation

\b m_t+\frac{c}{12}u_{xxx}+2mu_x+m_xu=0,\qquad m=u-u_{xx}\label
{CH_1}\e

\n can be written in Hamiltonian form

\b m_t=\{m,H_1\},\e where

\b \{F,G\}&\equiv& -\int \frac{\delta F}{\delta m}\Big(
\frac{c}{12}\partial^3+ m \partial+\partial\circ m
\Big)\frac{\delta G}{\delta m}dx \label{PB_1}\\
H_1&=&\frac{1}{2}\int mu {\text d}x.\e

\n The relation to the Virasoro algebra can be seen as follows
\cite{D03}. Suppose for simplicity that $m$ is $2\pi$ periodic in
$x$, i.e.

\b \label{eq7}
m(x)=\sum^{\infty}_{-\infty}L_{n}e^{inx}+\frac{c}{24},\e

\n (the reality of $m$ can be achieved by $L_{-n}=\bar{L}_n$) and
let us modify slightly (\ref{PB_1}) by a constant multiplier,

\b \{F,G\}&\equiv& -2\pi i \int_{0}^{2\pi} \frac{\delta F}{\delta
m}\Big( \frac{c}{12}\partial^3+ m \partial+\partial\circ m
\Big)\frac{\delta G}{\delta m}dx \label{PB_2}.\e

\n Then the Fourier coefficients $L_{n}$ close a classical
Virasoro algebra of central charge $c$ with respect to the Poisson
bracket (\ref{PB_2}):

\b \label{eq8}
\{L_{n},L_{m}\}=(n-m)L_{n+m}+\frac{c}{12}(n^{3}-n)\delta_{n+m,0}.\e

Further, we use the following form of the CH equation,

\b m_t+2\omega u_{x}+2mu_x+m_xu=0,\qquad m=u-u_{xx},\label{CH}\e

\n which can be obtained from (\ref{CH_1}) via $u\rightarrow
u+\frac{c}{12}$, and where apparently $\omega=c/8$.

Let us introduce the so-called independent conformal invariants of
the function $\phi=\phi(x,t)$:

\b \label{eq1}
p_{1} & = & \frac{\phi_{t}}{\phi_{x}}\nonumber \\
p_{2} & = & \{\phi;x\}\equiv\frac{\phi_{xxx}}{\phi_{x}}-\frac{3}{2}\frac{\phi_{xx}^{2}}{\phi_{x}^{2}}\nonumber \\
\e

\n Here $\{\phi;x\}$ denotes the Schwarz derivative. A quantity is
called conformally invariant if it is invariant under the M{\"o}bius
transformation

\b \label{eq2} \phi & \rightarrow&
\frac{\alpha\phi+\beta}{\gamma\phi+\delta}, \qquad
\alpha\delta\neq\beta\gamma. \e

\n For example, the KdV equation \b \label{eq3}
u_{t}+au_{xxx}+3uu_{x}=0
 \e
\n ($a$ is a constant) can be written in a Schwarzian form, i.e. in
terms of the conformal invariants (\ref{eq1}) as $p_{1}+ap_{2}=0$ or

\b \label{eq3b} \frac{\phi_{t}}{\phi_{x}}+a\{\phi;x\}=0 \e

\n where

\b \label{eq3a} u=a\{\phi;x\}. \e

The KdV and CH equations are also the geodesic flow equations for
the $L^{2}$ and $H^{1}$ metrics correspondingly on the
Bott-Virasoro group \cite{M98,CK03,K04,CK06,CKKT04,C00}.

\section{The Camassa-Holm equation in Schwarzian form}

It is known that the Camassa-Holm equation can be written as a
compatibility condition of the following two linear problems (Lax
pair) \cite{CH93}:

\b \label{eq10}
\Psi_{xx}&=&\Big(\frac{1}{4}+\lambda(m+\omega)\Big)\Psi,
 \\\label{eq11}
\Psi_{t}&=&\Big(\frac{1}{2\lambda}-u\Big)\Psi_{x}+\frac{u_{x}}{2}\Psi.
\e

\n In order to find the Schwarzian form for the CH equation we
proceed as follows. Let $\Psi_{1}$ and $\Psi_{2}$ be two linearly
independent solutions of the system (\ref{eq10}), (\ref{eq11}) and
let us define

\b \label{eq13} \phi=\frac{\Psi_{2}}{\Psi_{1}} \e

\n Then, from (\ref{eq11}) it follows that

\b \label{eq14} \frac{\phi_{t}}{\phi_{x}}=-u+\frac{1}{2\lambda} \e

\n According to the Theorem 10.1.1 from \cite{H76} due to
(\ref{eq10}) we also have \b \label{eq15} \{\phi; x\}=-2\lambda( m
+\omega) -\frac{1}{2} \e

\n From (\ref{eq14}), (\ref{eq15}) and the link between $m$ and
$u$ we obtain the Schwarz-Camassa-Holm (S-CH) equation:

\b \label{eq16}
(1-\partial^{2})\frac{\phi_{t}}{\phi_{x}}-\frac{1}{2\lambda}\{\phi;x\}=-\frac{3}{4\lambda}+\omega
\e

\n With a Galilean transformation, such that $\partial_t
\rightarrow \partial_t+b\partial_x$ with a suitable constant $b$,
one can absorb the constant on the right hand side and then the
S-CH equation (\ref{eq16}) acquires the form
$(1-\partial^{2})p_{1}+ap_{2}=0$ or

\b \label{eq17}
(1-\partial^{2})\frac{\phi_{t}}{\phi_{x}}+a\{\phi;x\}=0, \e

\n for some constant $a$. Applying the hodograph transform
$x\rightarrow \phi$, $t\rightarrow t$, $\phi \rightarrow x$ to the
S-CH (\ref{eq17}) and using the transformation properties of the
Schwarzian derivative \cite{H76}

\b \nonumber \{\phi;x\}=-\phi_{x}^{2}\{x;\phi\} \e

\n we obtain the following integrable deformation of the Harry Dym
equation for the variable $v=1/x_{\phi}$:

\b \label{eq17c}
v_{t}+v^{2}[v(v\partial_{\phi}^{-1}(v^{-1})_{t})_{\phi}]_{\phi}=av^{3}v_{\phi\phi\phi}
\nonumber \e

The conformal properties are preserved in some $2+1$ dimensional
generalizations. Indeed,  consider the equation \cite{CGP}

\b m_t+2\omega U_{xy}+2U_{xy}m+(U_y+\gamma) m_x=0, \qquad
m=U_x-U_{xxx},\e  where $\gamma$ is an arbitrary constant. This
equation reduces to CH equation in the case $x=y$ and
$u=U_x+\gamma$. The associated Lax pair is

\b \label{eqGP1}
\Psi_{xx}&=&\Big(\frac{1}{4}+\lambda(m+\omega)\Big)\Psi
 \\\label{eqGP2}
\Psi_{t}&=&\frac{1}{2\lambda}\Psi_y -(U_y+\gamma)
\Psi_{x}+\frac{U_{xy}}{2}\Psi. \e

\n In a similar manner this equation can be expressed in terms of
conformal invariants as \b \label{eqGPC}
(\partial-\partial^{3})\big(\frac{\phi_{y}}{\phi_{x}}-2\lambda
\frac{\phi_{t}}{\phi_{x}}\Big)+\partial_y\{\phi;x\}=0. \e

The equations (\ref{CH}) and (\ref{eq17}) with
$u=-\frac{\phi_{t}}{\phi_{x}}$ are not equivalent: as a matter of
fact (\ref{eq17}) implies (\ref{CH}), cf. \cite{W88}.  It is often
convenient to think that the Lax operator belongs to some Lie
algebra, and the corresponding eigenfunction - to the
corresponding group. Thus the relation between $u$ and $\phi$ (see
(\ref{eq13})) resembles the relation between the Lie group and the
corresponding Lie algebra, as pointed out in \cite{W88}. More
precisely, the following proposition holds:

Let $\phi$ be a solution of (\ref{eq16}). Then one can check
easily that $\Psi_{1}=\phi^{-1/2}_{x}$ and $\Psi_{2}=\phi
\phi^{-1/2}_{x}$ are two linearly independent solutions of
(\ref{eq10}). This is consistent with (\ref{eq13}). Therefore, the
general solution of (\ref{eq10}) is \b \label{eq18}
\Psi=\frac{A\phi+B}{\sqrt{\phi_{x}}} \e \n where $A$ and $B$ are
two arbitrary constants, not simultaneously zero.

Note that the expression (\ref{eq18}) is covariant with respect to
the M{\"o}bius transformation (\ref{eq2}), i.e. under (\ref{eq2}),
the expression (\ref{eq18}) transforms into an expression of the
same form but with constants

\b \label{eq19} A\rightarrow A'=\frac{\alpha A + \gamma
B}{\sqrt{\alpha\delta-\beta\gamma}},\qquad B\rightarrow
B'=\frac{\beta A + \delta B}{\sqrt{\alpha\delta-\beta\gamma}}. \e

\section{Other equations of the CH hierarchy}

Let us write the second equation of the CH Lax pair in the form

\b\Psi_{t}=-\mathcal{U}(x,\lambda)\Psi_{x}+\frac{1}{2}\mathcal{U}(x,\lambda)\Psi.\e

\n Taking $\mathcal{U}(x,\lambda)=\lambda v(x)$, the compatibility
condition gives $v=(m+\omega)^{-1/2}$ and the evolution equation

\b m_t+(\partial-\partial^3)(m+\omega)^{-1/2}=0.\e

\n Taking $\mathcal{U}(x,\lambda)=-\frac{1}{2\lambda}+u(x)+\lambda
v(x)$, we obtain the following integrable deformation of the CH
equation:

\b m_t+2\omega
u_{x}+2mu_x+m_xu+\alpha(\partial-\partial^3)(m+\omega)^{-1/2}=0,\e

\n where $m=u-u_{xx}$ and $\alpha$ is an arbitrary constant. (The
compatibility condition gives $v=2\alpha(m+\omega)^{-1/2}$ for an
arbitrary constant $\alpha$.)

An 'extension' of the CH hierarchy can be obtained if one
considers a more general Lax pair:

\b \label{L1} \Psi_{xx}&=&\mathcal{Q}(x,\lambda)\Psi,
 \\\label{L2}
\Psi_{t}&=&-\mathcal{U}(x,\lambda)\Psi_{x}+\frac{1}{2}\mathcal{U}_x(x,\lambda)\Psi,
\e

\n where

\b \label{L3} \mathcal{Q}(x,\lambda)&=&\lambda^n
q_n(x)+\lambda^{n-1} q_{n-1}(x)+\ldots+\lambda q_1(x)+\frac{1}{4},
 \\\label{L4}
\mathcal{U}(x,\lambda)&=&u_0(x)+\frac{u_1(x)}{\lambda}+\ldots
\frac{u_k(x)}{\lambda^k}. \e

\n The compatibility condition of (\ref{L1}), (\ref{L2}) gives the
equation

\b \label{L5} \mathcal{Q}_t=\frac{1}{2}
\mathcal{U}_{xxx}-2\mathcal{U}_x
\mathcal{Q}-\mathcal{U}\mathcal{Q}_x,\e

\n which, due to (\ref{L3}), (\ref{L4}), is equivalent to a chain
of $n$ evolution equations with $k+1$ differential constraints for
the $n+k+1$ variables $q_1$, $q_2$, $\ldots$, $q_n$, $u_0$, $u_1$,
$\ldots$, $u_k$ ($n$ and $k$ are arbitrary natural numbers, i.e.
positive integers):

\b  q_{n-r,t}&=&-\sum_{s=\max(0,r-k)} ^{r}(2u_{r-s,x}q_{n-s}+u_{r-s}q_{n-s,x}),\qquad r=0,1,\ldots,n-1, \nonumber \\
0&=&\frac{1}{2}(u_{r,xxx}-u_{r,x})-\sum_{s=1} ^{\min(n,k-r)}(2u_{r+s,x}q_{s}+u_{r+s}q_{s,x}), \nonumber \\
&\phantom{=}& \phantom{****************************}  r=0,1,\ldots,k-1, \nonumber\\
 0&=&\frac{1}{2}(u_{k,xxx}-u_{k,x}).\nonumber  \e

The two-component Camassa-Holm equation ($k=1$, $n=2$) was derived
earlier in \cite{SA}. More details and examples on the 'extended'
CH hierarchy can be found in \cite{I06}.

\section{Description of the whole CH hierarchy}

For the description of the whole CH hierarchy we need to introduce
the so-called recursion operator.

CH is a bi-hamiltonian equation, i.e. it admits two compatible
hamiltonian structures $J_1=(2\omega \partial +m\partial+\partial
m)$, $J_2=\partial-\partial^{3}$ :

\b m_t&=&-J_2\frac{\delta H_{2}[m]}{\delta m}=-J_1\frac{\delta
H_{1}[m]}{\delta m},\\ \label{H1a} H_1&=&\frac{1}{2}\int mu
\text{d}x,
\\\label{H2a} H_2&=&\frac{1}{2}\int (u^3+uu_x^2+2\omega u^2)
\text{d}x. \e

There exists an infinite sequence of conservation laws
(multi-Hamiltonian structure) $H_n[m]$, $n=0,\pm1, \pm2,\ldots$,
\cite{CH93,FS99,CGI2} such that \b J_2\frac{\delta
H_{n}[m]}{\delta m}&=&J_1\frac{\delta H_{n-1}[m]}{\delta m}.
\label{eq2a}\e

\n The recursion operator is $L\sim
J_2^{-1}J_1=(1-\partial^2)^{-1}[2(m+\omega)-\partial^{-1}m_x]\cdot$.

\n The eigenfunctions of the recursion operator are the squared
eigenfunctions of the CH spectral problem. More specifically, let
us for simplicity consider the case where $m$ is a Schwartz class
function, $\omega
>0$ and $m(x,0)+\omega > 0$. Then $m(x,t)+\omega
> 0$ for all $t$, e.g. see \cite{C01}. It is convenient to introduce the notation: $q\equiv m+\omega$.
Let $k^{2}=-\frac{1}{4}-\lambda \omega$, i.e. \b \label{lambda}
\lambda(k)= -\frac{1}{\omega}\Big( k^{2}+\frac{1}{4}\Big).\e

A basis in the space of solutions of (\ref{eq10}) can be
introduced: $f^+(x,k)$ and $\bar{f}^+(x,\bar{k})$. For all real
$k\neq 0$ it is fixed by its asymptotic when $x\rightarrow\infty$
\cite{C01}, see also \cite{ZMNP,CI06,CGI}: \b \label{eq6}
\lim_{x\to\infty }e^{-ikx} f^+(x,k)= 1, \e

\n   Another basis can be introduced, $f^-(x,k)$
and$\bar{f}^-(x,\bar{k})$ fixed by its asymptotic when
$x\rightarrow -\infty$ for all real $k\neq 0$: \b \label{eq6'}
\lim_{x\to -\infty }e^{ikx} f^-(x,k)= 1, \e

\n Since $m(x)$ and $\omega$ are real one gets that if $f^+(x,k)$
and $f^-(x,k)$ are solutions of (\ref{eq10}) then
\begin{equation}\label{eq:inv}
 \bar{f}^+(x,\bar{k}) = f^+(x,-k), \qquad \mbox{and} \qquad
 \bar{f}^-(x,\bar{k}) = f^-(x,-k),
\end{equation}
are also solutions of (\ref{eq10}). The squared solutions are
\begin{equation}\label{eq23} F^{\pm}(x,k)\equiv (f^{\pm}(x,k))^2, \qquad F_n^{\pm}(x)\equiv
F(x,i\kappa_n),
\end{equation}

\n where $F_n^{\pm}(x)$ are apparently related to the discrete
spectrum $k=i\kappa_n$, \b 0<\kappa_1<\ldots<\kappa_n<1/2.
\nonumber \e

\n Using the asymptotics (\ref{eq6}), (\ref{eq6'}) and the Lax
equation (\ref{eq10}) one can show that
\begin{equation}\label{eq23a} L_{\pm}F^{\pm}(x,k)=\frac{1}{\lambda}F^{\pm}(x,k).
\end{equation}

\n where

\b L_{\pm}=(\partial^2-1)^{-1}\Big[4q(x)-2\int_{\pm
\infty}^{x}\text{d}y \, m'(y)\Big]\label{eq44}\e

\n is the Recursion operator. The inverse of this operator is also
well defined.

If $\Omega(z)=\frac{P_1(z)}{P_2(z)}$ is a ratio of two polynomials
one can define $\Omega(L_{\pm})\equiv
P_1(L_{\pm})P_2^{-1}(L_{\pm})$ (provided $P_2(L_{\pm})$ is an
invertible operator). Then we can write the following nonlinear
evolution integro-differential (in general) equation

\b q_t+2q\tilde{u}_x+q_x\tilde{u}=0,\qquad
\tilde{u}=\frac{1}{2}\Omega(L_{\pm})\Big(\sqrt{\frac{\omega}{q}}-1\Big).\label{eq47}
\e




{\bf Example 1:} With $\Omega(z)=z$ one can  easily check that \b
\tilde{u}=\frac{1}{2}L_{\pm}\Big(\sqrt{\frac{\omega}{q}}-1\Big)=u\label{Lpm}\e
and thus the equation (\ref{eq47}) becomes the Camassa-Holm
equation (\ref{CH}) with Hamiltonian $H=H_1^{CH}=\frac{1}{2}\int
mu {\text d} x$.

{\bf Example 2:} $\Omega(z)=1/z$. The equation (\ref{eq47}) has
the form

\b
q_t+\frac{1}{4}\partial_x(\partial_x^2-1)\sqrt{\frac{\omega}{q}}=0,
\label{eq:Fokas} \e

\n i.e. the extended Dym equation \cite{CH93,FOR96,CGI2} with
Hamiltonian

\begin{equation}
H=\frac{1}{8}\int_{-\infty}^{\infty}\Big[\Big(\sqrt[4]{\frac{\omega}{q}}-\sqrt[4]{\frac{q}{\omega}}\Big)^2+\frac{\sqrt{\omega}q_x^2}{4q^{5/2}}
\Big]\text{d}x,\label{eq90}\end{equation}

\n which is, up to a constant, the (-1)-st Hamiltonian for the CH
equation,  $H_{-1}^{CH}$.

{\bf Example 3:} $\Omega(z)=z+\varepsilon/z$, where $\varepsilon$
is an arbitrary constant.

\b
q_t+2qu_x+q_xu+\frac{\varepsilon}{4}(\partial-\partial^3)q^{-1/2}=0,\e

The Hamiltonian of this equation is the first CH Hamiltonian with
an integrable perturbation, given by the (-1)-st CH Hamiltonian
(\ref{eq90}):

\b H&=&\frac{1}{2}\int_{-\infty}^{\infty} mu {\text
d}x+\frac{\varepsilon}{8}\int_{-\infty}^{\infty}\Big[\Big(\sqrt[4]{\frac{\omega}{q}}-\sqrt[4]{\frac{q}{\omega}}\Big)^2+\frac{\sqrt{\omega}q_x^2}{4q^{5/2}}
\Big]\text{d}x \nonumber \\ &=& H^{CH}_1+\varepsilon
H^{CH}_{-1}.\nonumber \e

Let us introduce the notation $\partial_{\pm}^{-1}\equiv \int_{\pm
\infty}^x{\text d}x$. The equations from the CH Hierarchy can be
written in the form \b
\frac{\partial_{\pm}^{-1}(\sqrt{q})_t}{\sqrt{q}}+\Omega(L_{\pm})\big(\sqrt{\frac{\omega}{q}}-1\Big)=0.\label{CHH}\e

The squared solutions (\ref{eq23}) form a complete basis in the
space of the Schwartz class functions $m(x)$, and $y$, $t$, can be
treated as some additional parameters. Also, the Generalised
Fourier Transform (GFT) for $q$ and its variation over this basis
is \cite{CGI2}

\begin{equation}\label{Compl1}
\sqrt{\frac{\omega}{q(x)}}-1= \pm\frac{1}{2\pi
i}\int_{-\infty}^{\infty}\frac{2k \mathcal{R}^{\pm}(k)} {\omega
\lambda(k)} F^{\pm}(x,k)\text{d}k +
\sum_{n=1}^{N}\frac{2\kappa_n}{\omega
\lambda_n}R_n^{\pm}F_n^{\pm}(x),
\end{equation}
\b \frac{\partial_{\pm}^{-1}\delta(\sqrt{q})}{\sqrt{q}}&=&
\frac{1}{2\pi
i}\int_{-\infty}^{\infty}\frac{i\mathcal{R}^{\pm}(k)}{\omega
\lambda(k)}\delta F^{\pm}(x,k)\text{d}k \nonumber  \\
& \pm & \sum_{n=1}^{N}\Big[\frac{\delta R_n^{\pm}-R_n^{\pm}\delta
\lambda_n }{\omega \lambda_n}F_n^{\pm}(x)
+\frac{R_n^{\pm}}{i\omega\lambda_n}\delta \kappa_n
\dot{F}_n^{\pm}(x) \Big]. \label{Compl2}  \e

\n Here $\dot{F}_n^{\pm}(x)\equiv \frac{\partial}{\partial
k}F^{\pm}(x,k)|_{k=i\kappa_n}$. The generalized Fourier
coefficients $\mathcal{R}^{\pm}(k)$, $R_n^{\pm}$, together with
the set of discrete eigenvalues, are called scattering data. The
variation is with respect to any additional parameter, e.g. $y$,
$t$.  Due to the completeness of squared eigenfunctions basis,
from (\ref{CHH}), (\ref{Compl1}) and (\ref{Compl2}) we have linear
differential equations for the scattering data: \b
\mathcal{R}^{\pm}_t \mp
ik\Omega(\lambda^{-1})\mathcal{R}^{\pm}(k)=0, \\ R^{\pm}_{n,t}
\pm \kappa_n \Omega(\lambda_n^{-1}) R^{\pm}_n=0, \\
\lambda_{n,t}=0. \e


\n The GFT for other integrable systems is derived e.g. in
\cite{K76,GeHr1,G86,GI92,GY94,IKK94}.

{\bf Example 4:} Consider again the two dimensional CH
generalisation \b q_t+2U_{xy}q+(U_y+\gamma)q_x=0, \qquad
q=U_x-U_{xxx}+\omega,\e with arbitrary constants $\omega$ and
$\gamma$. This equation can be written as \b
(\sqrt{q})_t+[(U_y+\gamma)\sqrt{q}]_x=0.\e  Then

\b \partial_{\pm}^{-1}(\sqrt{q})_t+(U_y+\gamma)\sqrt{q}+\beta=0,\e
where $\beta$ is an integration constant. Further, with the choice
$\beta=-\gamma\sqrt{\omega}$ and due to the identity \b
U_y=-\frac{1}{2}L_{\pm}
\Big(\frac{\partial_{\pm}^{-1}(\sqrt{q})_y}{\sqrt{q}}\Big), \e

the equation can be written in the form \b
\frac{\partial_{\pm}^{-1}(\sqrt{q})_t}{\sqrt{q}}-
\frac{1}{2}L_{\pm}
\Big(\frac{\partial_{\pm}^{-1}(\sqrt{q})_y}{\sqrt{q}}\Big)-\gamma\big(\sqrt{\frac{\omega}{q}}-1\Big)=0.
\label{2d}\e

Again, from  (\ref{2d}), (\ref{Compl1}) and (\ref{Compl2}),
considering variations with respect to $y$ and $t$ we obtain
linear equations for the scattering data:

\b \mathcal{R}^{\pm}_{t}
-\frac{1}{2\lambda}\mathcal{R}^{\pm}_{y}\pm 2ik\gamma
\mathcal{R}^{\pm}=0,
\\R^{\pm}_{n,t}
-\frac{1}{2\lambda_n}R^{\pm}_{n,y}\mp2\gamma\kappa_n R^{\pm}_n=0.
\e

\n E.g. when $\gamma=0$ the solution is any function (with
appropriate decaying properties) of $t+2\lambda y$:

\b \mathcal{R}^{\pm}(y,t)=\mathcal{R}^{\pm}(t+2\lambda y), \qquad
R^{\pm}_{n}(y,t)=R^{\pm}_{n}(t+2\lambda_ny). \e

Other possible choices for $\Omega(z)$  (\ref{eq47}) produce the
other members of the Camassa-Holm hierarchy.

\section{Inverse scattering transform}\label{sec:10}

Inverse scattering method for the hierarchy (\ref{eq47}) is the
same as the one for the CH equation \cite{CGI}. The only
difference is the time-dependence of the scattering data (and/or
the $y$-dependence, etc). For example, the inverse scattering is
simplified in the important case of the so-called reflectionless
potentials, when the scattering data is confined to the discrete
spectrum. This class of potentials corresponds to the $N$-soliton
solutions of the CH hierarchy. In this case the time evolution of
the scattering data is $R_n^+$ is  \b
R_n^+(t)=R_n^+(0)\exp\Big(-\kappa_n \Omega(\lambda_n^{-1})t
\Big).\e

The $N$-soliton solution is \cite{CGI}
\begin{equation}
q(x,t)= \int_0^{\infty} \delta(x-g(\xi,t)) p(x,t) \text{d}\xi,
\label{q}\end{equation} where $g(\xi,t)$ can be expressed through
the scattering data as
\begin{eqnarray} g(\xi,t)&\equiv&\ln
\int_0^{\xi}\Big(1-\sum_{n,p}\frac{R_n^+(t)
\underline{\xi}^{-2\kappa_n}}{\kappa_n+1/2}A^{-1}_{np}
[\underline{\xi},t]\Big)^{-2}\text{d} \underline{\xi},\label{g}\e
with \b  A_{pn}[\xi,t]&\equiv&
\delta_{pn}+\frac{R_n^+(t)\xi^{-2\kappa_n}}{\kappa_p+\kappa_n}
\nonumber
\end{eqnarray}

\n and
\begin{equation}
p(\xi,t)=\omega \xi^{-2}g_{\xi}^{-1}(\xi,t).
\label{p}\end{equation} In particular, for the CH equation $
q_t+uq_x=-2qu_x$, from (\ref{q}) it follows
\begin{equation}
 \dot{g}(\xi,t)=\frac{1}{2}\int_0^{\infty}
e^{-|g(\xi,t)-g(\underline{\xi},t)|}p(\underline{\xi},t)d\underline{\xi}
 -\omega, \qquad
\dot{g}(\xi,t)=u(g(\xi,t),t), \nonumber\end{equation} therefore
$g(x,t)$ in (\ref{g}) is the diffeomorphism (Virasoro group
element) in the purely solitonic case \cite{CI07}. The situation
when the condition $q(x,0)\equiv m(x,0)+\omega>0$ on the initial
data does not hold is more complicated and requires separate
analysis \cite{K05} (if $m(x,0)+\omega$ changes sign there are
infinitely many positive eigenvalues accumulating at infinity and
singularities might appear in finite time \cite{CE98,C00,C01}).

The explicit construction of the peakon solutions ($\omega=0$) is
also known \cite{CH93,BBS98,BBS99}, e.g. a single peakon
travelling with speed $c$ is $u_c(x,t)=ce^{-|x-ct|}$.  The peakons
are the only solitary waves if $\omega=0$, cf. \cite{L05}. They
have to be interpretted as weak solutions due to the fact that
they are not continuously differentiable - e.g. see \cite{BC07}.
The peakons however interact like solitons \cite{CH93,BBS99}. Some
nonintegrable generalizations of the CH equation also have been
studied recently, e.g. \cite{SS07}.

\section{Acknowledgements}

The author acknowledges funding from the Irish Research Council
for Science, Engineering and Technology. The author is thankful to
Prof. Adrian Constantin and Prof. Vladimir Gerdjikov for many
stimulating discussions and to an anonymous referee for important
comments and suggestions.

\end{document}